\documentstyle[11pt]{article}
\begin{document}
\renewcommand{\thepage}{ }
\begin{titlepage}
\title{
\hfill
\vspace{1.5cm}
{\center \bf $\nu=1/2$ quantum Hall effect in the Aharonov-Casher geometry
in a mesoscopic ring}
\author{R. M\'elin\thanks{Present Address:
International School for Advanced
Studies (SISSA), Via Beirut 2--4, 34014 Trieste, Italy;
e--mail: melin@crtbt.polycnrs--gre.fr}
 and B. Dou\c{c}ot\thanks{Present address: LPTHE, Jussieu,
tour 24-14, 4 place Jussieu, 75252 Paris Cedex France;
e--mail: doucot@lpthe.jussieu.fr}
\\
{}\\
CRTBT-CNRS, 38042 Grenoble BP 166X c\'edex France}}
\date{}
\maketitle
\begin{abstract}
\normalsize
\noindent
We study the effect of an electric charge in the middle of a ring
of electrons in a magnetic field such as $\nu = 1/2$. In the
absence of the central charge, a residual current should appear
due to an Aharanov-Bohm effect. As the charge varies, periodic currents
should appear in the ring. We evaluate the amplitude of these
currents, as well as their period as the central charge varies.
The presence of these currents should be a direct signature of the
existence of a statistical gauge field in the $\nu=1/2$ quantum Hall effect.
Numerical diagonalizations for a small number of electrons on the sphere
are also carried out. The numerical results up to
9 electrons are qualitatively consistent with the
mean field picture.

\end{abstract}
\end{titlepage}

\newpage
\renewcommand{\thepage}{\arabic{page}}
\setcounter{page}{1}
\baselineskip=17pt plus 0.2pt minus 0.1pt
\newpage

\section{Introduction}
Recently, new experimental and theoretical developments have generated
alternative viewpoints in the current understanding
of the fractional quantum Hall effect.
It seems the key idea is the notion of composite fermions,
first introduced by Jain \cite{Ref1}, which establishes a correspondence
between the fractional quantum Hall effect at lowest Landau level
filling $\nu=p/(2 m p \pm 1)$ and the integer quantum Hall effect
with $p$ filled Landau levels.
The mapping is achieved by attaching $2 m$ flux quanta to each electron,
which preserves fermionic statistics, and by first treating the
statistical fluxes at the mean field level.
This description has received considerable attention especially in the
vicinity of $\nu=1/2$.
Experimentally, the system is characterized by a vanishing energy gap,
and various physical properties strongly suggest the presence of a new
Fermi liquid \cite{Ref2} in this problem. On the theoretical side,
the composite picture with $m=1$ has been advocated by Halperin,
Lee and Read as a powerful microscopic basis to understand these
Fermi liquid-like properties in the vicinity of $\nu=1/2$ \cite{Ref3}.
The vanishing of the gap for the sequence of states at $\nu=p/(2p \pm 1)$
as $p$ goes to infinity receives for instance a very simple
interpretation in this framework since the
composite fermions experience an average flux equal to $\phi_0/p$
per particle, so the corresponding Landau level spacing goes as $1/p$.
If the statistical gauge field fluctuations are taken into account,
some complications arise since a logarithmic divergence of the
effective mass as $p \rightarrow + \infty$ is predicted \cite{Ref3}
\cite{Ref4}. There has been some trends in recent observation suggesting
an effective mass enhancement \cite{Ref5} but this question is not
settled yet.
This theory has also been tested by numerical diagonalizations on finite
systems \cite{Ref6} \cite{Ref7}, and the composite fermion picture
has been shown to predict for instance the right quantum numbers for
the ground state and low-lying states \cite{Ref6}, as well as the
scaling of the ground state energy versus particle number \cite{Ref7}.
Furthermore, a trial wave function inspired by these considerations
provides a very good understanding of ground state correlations. We note
however that this wave function is not merely the singular gauge
transformation applied to the Slater determinant of composite fermions
moving in zero external field, but it has to be improved by the combined
effect of a short range Jastrow factor and a global projection
on the lowest Landau level.

In this paper, we suggest a different test for the composite fermion approach.
The basic idea is connected to the fact that if an external electric
field generates a spatially dependent electronic density, the
average flux acting on the composite fermions is no longer vanishing
everywhere. We wish to detect such a variation of the effective flux.
In the experiment we propose, local density fluctuations
are created in a ring of $\nu = 1/2$ fermions, thanks to the presence of
an electric charge localized in the middle of the ring. Due to the presence
of this charge, the electrons have a trend to accumulate at the external edge
of the ring if the central charge is negative,
and to be repelled from the inner edge, which creates a non zero
effective flux through the ring. A consequence of this
non zero flux is the existence
of observable persistent currents which
vary periodically with the value of the charge
in the center of the ring.
At first glance, this phenomenon is reminiscent of the Aharonov-Casher
effect, which is expected to take place if a flux tube moves around a
fixed charge. \cite{Ref8}. This effect is a direct consequence of the
Aharonov-Bohm effect and of Lorentz invariance, since the static electric
field generated by the charge induces a non-vanishing magnetic field
in the flux-tube co-moving frame. Experimentally, it has been observed with
superconducting vortices in Josephson junction arrays \cite{Ref9}.
However, we should emphasize here that the composite fermions do not
carry the physical magnetic field which obeys Maxwell's
equations, but rather a Chern-Simon flux, so the two situations are
physically quite different.

This paper is organized as follows. We first sum up the formalism to deal with
the mean field approach of the $\nu = 1/2$ quantum Hall effect. We then
describe the geometry of the experiment, and propose a simplified geometry
to get rid of the effects of curvature. In the absence of the central
charge, one is able to solve exactly the eigenstates of the problem in the
geometry of the experiment. In the presence of a central charge, the ring
is electrically polarized. We propose a self consistent approach to find
the electron density in the ring. Due to the presence of screening, the
density fluctuations are localized in the vicinity of the edges of the ring.
We finally obtain an order of magnitude for the average magnetic field
through the ring. We calculate the current in the ring as
a function of the central charge.
We then present numerical computations of the effect on the sphere. After
briefly writing the mean field theory on the sphere, we give numerical
results for $\nu=1/2$.
A conclusion discusses the various results and mentions some open
questions.

\section{Mean field theory at $\nu=1/2$}
\subsection{Gauge transformation at $\nu=1/2$}
We first sum up the mean field theory treatment of the half field
Landau level \cite{Ref3}.
It is possible to transform the original fermions in the magnetic field
into new fermionic composite particles, with two flux quanta attached to
them. The reason why attaching two flux quanta leads to fermions is that when
one interchanges two particles with a flux $\phi$ attached to them, one gets
a phase factor of $\exp{i \theta}$ in the wave function, with
\begin{equation}
\theta = \pi (1 + \frac{\phi}{\phi_0})
,
\end{equation}
where $\theta = \pi$ refers to fermionic particles. $\phi_0 = h/e$ is the
flux quantum.
The case $\phi = 2 \phi_0$ thus corresponds to fermionic particles since
in this case $\theta$ is $\pi$ modulo $2 \pi$.
The flux tubes are fixed via a non local gauge transformation, implemented
in the following way.
The second quantized form of the kinetic term in the Hamiltonian reads
\begin{equation}
\hat{K} = \frac{1}{2 M} \int d^{2} {\bf x} \psi^{+}({\bf x})
(- i \hbar {\bf \nabla} + e {\bf A}(x))^{2} \psi({\bf x})
,
\end{equation}
where $M$ is the band electronic mass, $e$ the absolute value of the
electronic charge and $\psi^{+}({\bf x})$ the electronic field.
The gauge transformation which realizes the passage from the electronic
field $\psi^{+}({\bf x})$ to the composite fermionic field $\psi_C^{+}({\bf x})$
is given by
\begin{equation}
\psi_C^{+}({\bf x}) = \psi^{+}({\bf x})
\exp{\{ - 2 i \int d^{2} {\bf x}' arg({\bf x}-{\bf x}') \hat{\rho}({\bf x}')\}}
\label{eq1}
\end{equation}
where
\begin{equation}
\hat{\rho}({\bf x}) = \psi^{+}({\bf x}) \psi({\bf x})
= \psi_C^{+}({\bf x}) \psi_C({\bf x}) = \hat{\rho}_C({\bf x})
,
\end{equation}
and $arg({\bf x}-{\bf x}')$ is the angle between the vector
${\bf x} - {\bf x}'$ and the $x$ axis.
The factor $2$ in the exponential of (\ref{eq1}) stands for two flux
tubes
attached to the electrons to form the composite fermions. In terms of the
composite fermions, the kinetic energy reads
\begin{equation}
\hat{K} = \frac{\hbar^{2}}{2 M} \int d{\bf x} \psi_C^{+}({\bf x})
(- i {\bf \nabla} + \frac{e}{\hbar} {\bf A}({\bf x}) - {\bf a}({\bf x}))^{2}
\psi_C({\bf x})
,
\end{equation}
where the statistical gauge field is
\begin{equation}
{\bf a}(x) = 2 \int d{\bf x}' \frac{\hat{{\bf z}}\wedge({\bf x} -{\bf x'})}
{|{\bf x}-{\bf x}'|^{2}} \hat{\rho}_C({\bf x}')
.
\end{equation}
The interaction term is
\begin{equation}
\hat{V} = \frac{1}{2} \int d{\bf x} d{\bf x}' v({\bf x}-{\bf x}')
:\hat{\rho}({\bf x}) \hat{\rho}({\bf x}'):\\
= \frac{1}{2} \int d{\bf x} d{\bf x}' v({\bf x}-{\bf x}')
:\hat{\rho}_C({\bf x}) \hat{\rho}_C({\bf x}'):
,
\end{equation}
$v({\bf x})$ being the Coulomb interaction
\begin{equation}
\label{eq13}
v({\bf x}) = \frac{e^{2}}{4 \pi \epsilon_0 \epsilon_r |{\bf x}|}
,
\end{equation}
where $\epsilon_r$ is the relative dielectric constant of the material.
For GaAs, $\epsilon_r = 12.6$.

\subsection{Mean field theory}
In the mean field theory approach, the density of fermions $\rho({\bf x})$ is
assumed to be constant. Since the average density $n_0$ is related to
the field by the condition of half filling
\begin{equation}
n_0 = \frac{B}{2 \phi_0}
\label{eq2}
,
\end{equation}
the mean field value of the statistical flux exactly cancels the external
magnetic flux, and gives a zero residual magnetic field.
The Hamiltonian of the system of composite fermions is simply the
Hamiltonian of a collection of free fermions, but with a renormalized
effective mass $M^{*}$
\begin{equation}
\hat{H}_{M.F.} = \frac{1}{2 M^{*}} \int \psi^{+}({\bf x}) (-i \hbar
{\bf \nabla})^{2} \psi({\bf x}) d{\bf x} + \hat{V}
.
\end{equation}
If $B=10 T$, the
value of the effective mass for GaAs is $M^{*} \simeq 4 M \simeq 0.27 M_e$,
where $M_e$ is the bare electronic mass. $M^{*}$
increases as the square root of the magnetic field
for larger magnetic fields \cite{Ref3}.
If the density $n({\bf x})$ deviates from the average density $n_0$,
then a residual effective magnetic field appears, equal to
\begin{equation}
B - 2 \phi_0 n({\bf x}) = 2 \phi_0 (n_0 - n({\bf x}))
= - 2 \phi_0 \delta n(~{\bf x})
.
\end{equation}

\section{Geometry of the experiment}
The electrons are confined on a small two--dimensional ring.
The ring is the set of points ${\bf x}$, such as $r_0 < |{\bf x}| < r_0 + L$.
The dimensions of the ring are of the order $r_0 \simeq 1 \mu m$
and $L \simeq 0.1 \mu m$. We look for the response of the system to an
extra charge $Q$ added in the center of the ring.
As we shall see in the next section, one is able
to solve exactly the Schr\"odinger equation of the electrons on the ring
in the absence of the central charge and in the absence of
Coulomb interaction between electrons.
The wave functions are given in terms of Bessel functions. In order to
simplify the treatment of the problem, we change the geometry.
(see figure \ref{Fig1}).
Instead of a ring, we shall use a rectangle of size $L$ and $R = 2 \pi r_0$.
The $x$ axis is chosen along the $R$ side of the rectangle, and the $y$
axis along the $L$ side of the rectangle.
We impose cyclic boundary conditions
in the $x$ direction, which means that $\psi(x+R,y) = \psi(x,y)$ for the
wave functions. The spectrum and the wave functions are much more simpler
for the approximate geometry. Two types of potentials
in the $x$ direction will be investigated: an infinite square
well and also a parabolic well, to understand
the effects of a gradual fall--off of the electronic
density as may be the case in experiments.

\section{$\nu = 1/2$ electrons on the ring in the
absence of the central charge}
One is able to solve exactly the eigenvalue problem in the absence of
the central charge, in the absence of a Coulomb interaction
between electrons and in the ring geometry,
as well as in the simplified geometry.
The magnetic field is related to the density by the condition of half filling
(\ref{eq2}), so that the total effective magnetic field
through the ring is zero,
and the magnetic field in the hole inside the ring is uniform and given
by
\begin{equation}
B = 2 \phi_0 n_0
.
\end{equation}
The number of electrons $N$ on the ring is given by
\begin{equation}
N = \frac{ \pi B L (L+2 r_0)}{2 \phi_0}
\label{eq3}
.
\end{equation}
If we take $B = 20 T$, the number of electrons on the ring is $N=1494$.
These $N$ composite fermions feel a zero magnetic
field, but feel the vector potential
created by the flux tube in the hole of the ring. This is a typical
Aharanov-Bohm situation \cite{ref2} and one expects the presence of
electronic currents which are periodic
in $\varphi/ \phi_0$, where $\varphi$ denotes
the flux in the center of the ring. Its value is simply
\begin{equation}
\frac{\varphi}{\phi_0} = \frac{\pi r_0^{2} B}{\phi_0}
.
\end{equation}
Similar situations have already
been analyzed in different geometries \cite{Ref10} \cite{Ref11}.
The gauge is chosen such as ${\bf A}({\bf x})=A(|{\bf x}|)
{\bf e_{\theta}}$, with
\begin{equation}
A(r) = \frac{\varphi}{2 \pi r}
\end{equation}
if $r>r_0$. One is left with a problem of free electrons in the vector
potential ${\bf A}(r)$ with a Hamiltonian
\begin{equation}
\hat{H} = \frac{1}{2 M^{*}} (- i \hbar {\bf \nabla} + e {\bf A})^{2}
.
\end{equation}
Because of the rotational invariance of the problem, the wave functions
can be chosen with a definite orbital momentum $m \hbar$
\begin{equation}
\psi(r,\theta) = e^{i m \theta} \chi(r)
.
\end{equation}
The Schr\"odinger equation for $\chi(r)$ reads
\begin{equation}
\chi''(r)+\frac{\chi(r)}{r}+(k^{2}-\frac{1}{r^{2}}
(m+\frac{\varphi}{\phi_0})^{2}
)\chi(r) = 0
\label{eq4}
,
\end{equation}
which is a Bessel equation where we have set $E = \hbar^{2}
k^{2}/2M^{*}$. The general solution of this equation is
\begin{equation}
\chi(r) = A J_{|m+\frac{\varphi}{\phi_0}|}(kr)
+ B Y_{|m+\frac{\varphi}{\phi_0}|}(kr)
.
\end{equation}
The coefficients $A$ and $B$ are determined by the normalisation of the
wave function and by the fact that the wave function vanishes at the edges
of the sample, so that $\chi(r_0)=\chi(r_0+L)=0$. The wave vector $k$ is
found to be the solution of
\begin{equation}
J_{|m+\frac{\varphi}{\phi_0}|}(k r_0) Y_{|m+\frac{\varphi}{\phi_0}|}(k(r_0+L))
- Y_{|m+\frac{\varphi}{\phi_0}|}(k r_0) J_{|m+\frac{\varphi}{\phi_0}|}(k(r_0+L))
=0
.
\end{equation}
In order to obtain simpler equations for the energy levels and the
wave functions, we give the form of the solution in the simplified geometry.
We first treat the case $\varphi = 0$. In this case, the problem simply
corresponds to free electrons on a rectangle. The wave functions are
given by
\begin{equation}
\psi_{m,n} (x,y) = \sqrt{\frac{2}{RL}} e^{-i m \frac{2 \pi}{R} x}
\sin{\left( n \frac{\pi}{L}y \right)}
\label{eq27}
,
\end{equation}
and the energy levels are
\begin{equation}
E(m,n) = \frac{\hbar^{2}}{2 M^{*}}\left(
\left( \frac{2 \pi}{R} \right)^{2} m^{2}
+ \left(\frac{\pi}{L}\right)^{2} n^{2}\right)
.
\end{equation}
The electrons belong to a Fermi sea, and the Fermi wave vector $k_F$ is
approximately determinated by the relation
\begin{equation}
k_F = \sqrt{\frac{4 \pi N}{L R}}
,
\end{equation}
where the number of fermions $N$ is given by (\ref{eq3}).
Since $L \ll R$, the wave vector increment $2 \pi/R$ in the $k_x$
direction is much smaller than the increment $\pi/L$ in the $k_y$ direction,
so that the Fermi sea can be viewed as a collection of channels labeled
by the integer $n > 0$.
The Fermi sea is drawn on figure \ref{Fig2}.
In the case $B = 20 T$, the Fermi sea is made
up of six channels.
In the absence of a magnetic flux $\varphi$ through the hole of the ring,
or if the magnetic flux is a multiple of the flux quantum $\phi_0$,
one can compute the electronic density of the Slater determinant
\begin{equation}
|\psi_0 \rangle = \prod_{m,n \in F.S.} \psi_{m,n}^{+} | 0 \rangle
\label{eq23}
,
\end{equation}
where the fermionic quantum numbers $m$ and $n$ belong to the Fermi sea of 
figure \ref{Fig2} and $|0\rangle$ is the vacuum.
The density profile is plotted on figure \ref{Fig3}.
As expected, the density is zero on the edges and the density profile exhibits
Friedel oscillations. The non uniformity of the density induces a non uniform
electrostatic field, which modifies the one-particle states. Thus, due
to finite size effects, (\ref{eq23}) is not the true ground state
for the mean field approximation, whereas
it would be the true ground state on an infinite plane.

What happens if the flux $\varphi$ in the hole of the ring is non zero?
Let us first consider the case where $\varphi/\phi_0$ is an integer.
As we see from equation (\ref{eq4}), we can deduce the physics from the
case $\varphi=0$ by replacing $m$ by $m+\varphi/\phi_0$. The Fermi sea is
translated in the reciprocal space by a factor $\Delta k_x = 2 \pi
\varphi/ R \phi_0$ along the $k_x$ direction.

If $\varphi/\phi_0$ is not an integer, the situation is somewhat different.
One expects in this case the appearance of a current due to an Aharanov-Bohm
effect \cite{Ref12}.
We note $\varphi/\phi_0 = \Delta m + \delta m$, with $\Delta m$ an
integer and $\delta m$ a real number, such as $|\delta m| < 1/2$.
The energy $E(m,n)$ reads
\begin{equation}
E(m,n) = \frac{\hbar^{2}}{2 M^{*}}
[ (\frac{2 \pi}{R})^{2} (m + \Delta m + \delta m)^{2}
+ (\frac{\pi}{L})^{2} n^{2}]
.
\end{equation}
The Fermi sea is translated by the wave vector
$\Delta k_x =- 2 \pi \Delta m / R$, so that the wave functions
become
\begin{equation}
\psi_{m,n}(x,y)=\sqrt{\frac{2}{RL}} e^{i(m-\Delta m)\frac{2 \pi}{R}x}
\sin{n \frac{\pi}{L}y}
\end{equation}
The Fermi sea becomes unstable as $|m| = 1/2$.
To see this, consider $2 N_0 + 1$ fermions in a given channel $n$.
If $|\delta m|$ is less than $1/2$, the fermions have the possibility
to have their orbital quantum numbers in the interval
$[-\Delta m - N_0, -\Delta m + N_0]$, or in the interval
$[-\Delta m - N_0 \pm 1, -\Delta m + N_0 \pm 1]$.
The condition for the latter configuration to be stable is that it has
a lower energy than the former, namely
\begin{equation}
\frac{\hbar^{2}}{2 M^{*}} (\frac{2 \pi}{R})^{2}
\sum_{m=-\Delta m - N_0}^{-\Delta m + N_0} (m + \Delta m + \delta m)^{2}
<
\frac{\hbar^{2}}{2 M^{*}} (\frac{2 \pi}{R})^{2}
\sum_{m=-\Delta m - N_0 \pm 1}^{-\Delta m + N_0 \pm 1}
(m + \Delta m + \delta m)^{2}
,
\end{equation}
which is satisfied for $|\delta m| < 1/2$, independently on $N_0$.
The Fermi sea of $2 N_0 +1$ fermions in the channel $n$ is thus in the interval
$m \in [\Delta m - N_0 , \Delta m + N_0 ]$ if $-1/2 < \delta m
< 1/2$.
What is the Aharonov-Bohm current in the ring for a non integer value of
$\varphi/\phi_0$ ? The quantum mechanical, gauge invariant current operator
reads
\begin{equation}
{\bf j} ({\bf x}) = \frac{- i \hbar}{2 M^{*}}(\overline{\psi} {\bf \nabla}
\psi - \psi {\bf \nabla} \overline{\psi}) + \frac{e}{M^{*}}
{\bf A} |\psi|^{2}
.
\label{eq17}
\end{equation}
In the simplified geometry, the current has only a component parallel to
the $x$ axis. The contribution of the first term in (\ref{eq17}) is
\begin{equation}
\frac{\hbar}{M^{*}} \frac{2 \pi}{R} (m-\Delta m)
,
\end{equation}
and the contribution of the term proportional to the vector potential reads
\begin{equation}
\frac{\hbar}{M^{*}} \frac{2 \pi}{R} (\Delta m + \delta m)
.
\end{equation}
The total current is thus
\begin{equation}
j_x(m,n) = \frac{\hbar}{M^{*}} \frac{2 \pi}{R}(m + \delta m)
|\psi_{m,n}|^{2}
.
\end{equation}
After a summation over the Fermi sea, we obtain the total current
\begin{equation}
J_x(y) = \sum_{(m,n)\in F.S.} j_x(m,n) = \frac{2 \hbar}{M^{*}}\frac{2 \pi}{R} \delta m
\sum_{(m,n) \in F.S.} |\psi_{m,n}|^{2}
,
\end{equation}
If $\delta m$ = 0, the flux in the hole of the ring is a multiple of the
flux quantum, and everything happens as if the fermions would not see
the flux. If $\delta m \ne 0$, a current exists. The intensity of the current
is given by
\begin{equation}
I = - e \int_{0}^{L} J_x(y) dy
= - \frac{2 \pi e \hbar N}{M ^{*} R^{2}} \delta m
.
\end{equation}
The maximum value of the current is
\begin{equation}
I_{max} = \frac{e h N}{2 M^{*} R^{2}}
\label{eq16}
.
\end{equation}
The numerical value of $I_{max}$ is $5.8 nA$ for $B=20T$.

\section{$\nu = 1/2$ electrons on a ring in the presence of the external charge}
In the presence of a positive central charge, the electrons are expected
to accumulate near the internal edge of the ring and a depletion of electrons
is expected at the external edge, leading to a charge transfer from the
external to the internal edge. This charge transfer induces a deviation from
the $\nu=1/2$ value and, if ones applies the ideas of the mean field theory,
the total effective flux, due to the external magnetic field plus the
statistical gauge field is no longer zero, so that an effective flux penetrates
through the sample, creating an Aharanov-Bohm current which is periodic as
a function of the central charge.

The first step is to evaluate the screening of the central charge by the
electron gas on the ring. Many levels of approximation are possible.
The most accurate approximation is to take into account that the one particle
Slater determinant (\ref{eq23}) is not the true ground state of the fermions,
due to Friedel oscillations induced by the edges.
This can be implemented in a recursive way
by starting from the state (\ref{eq23}), computing the electronic density,
deducing from it the electrostatic field,
and reiterating, that is to compute the corrections to the one-particle
states in the presence of the Friedel oscillations, and re-compute
the electronic density. The effect of the charge is then treated at
the linear order, with the full response function.
Within these approximations, one expects the density of electrons to increase
near the internal edge of the ring, and to decrease near the external
edge.
Moreover, this average density profile should be modulated by Friedel
oscillations as well as oscillations at the Thomas-Fermi wave vector.
In our case, as we
shall see, the Thomas-Fermi wave vector is greater than the Fermi wave vector.
We shall not use this refined
approximation scheme because of its computational complexity.
Since we focus only on the charge transfer from one
edge of the ring to the other,
and not on the details of the variations of the density profile,
we do not take into account the existence of
Friedel oscillations. We take the state (\ref{eq23}) as an approximate
ground state and we look for the effects of the screening of the central charge.
In other words, we hope the density-density response functions are not
very different if we replace the true Hartree-Fock state by the approximate
one given by (\ref{eq23}).

At this stage, two approximations are possible. The most refined one consists
in taking into account the full response function of the fermions. This shall
be done in section 5.4. A more approximate treatment is to treat the
screening in the Thomas-Fermi approximation, which is the aim of section
5.3. As we shall see, the two approximations lead to similar results
as far as averaged quantities are concerned, namely the average effective
flux
penetrating through the ring due to the presence of the central charge.

\subsection{Screening in two dimensions}
As an introduction to the problem of screening in a finite geometry, we treat
the case of the screening of a single charge in an infinite
two--dimensional gas
of electrons. The solution of this problem will lead to the expression of
the Thomas-Fermi wave vector $q_{T F}$ in two dimensions. The case of the
screening of a charge in three dimensions is treated in reference \cite{Ref13}
with the Thomas-Fermi approximation. We do nothing but transpose the
argument of \cite{Ref13} to the case of an electron gas constrained
on a two--dimensional layer, with three dimensional Coulomb interactions.
The potential created by the external point charge located at the origin is
noted $\phi^{ext}$. We note $\rho^{ind}$ the variation of density induced
by the presence of the extra charge in the gas of electrons, plus the
uniform background of positive charges. $\phi$ is the total potential
created by the extra charge, the electrons and the background of positive
charges. Since the two--dimensional Fourier transform of $1/|{\bf x}|$ is
$2 \pi/|{\bf q}|$, we have
\begin{equation}
\phi({\bf q}) - \phi^{ext}({\bf q}) =
 \frac{\rho^{ind}({\bf q})}{2 \epsilon_0 \epsilon_r |{\bf q}|} 
.
\end{equation}
The dielectric constant is defined by
$\phi^{ext}({\bf q}) = \epsilon({\bf q}) \phi({\bf q})$,
and we assume that
$\rho^{ind}({\bf q}) = \chi({\bf q}) \phi({\bf q})$.
We thus obtain
\begin{equation}
\epsilon({\bf q}) = 1 - \frac{\chi({\bf q})}{2 \epsilon_0
\epsilon_r |{\bf q}|}
.
\end{equation}
If the total potential is a slowly varying function of the position,
one can define
\begin{equation}
\epsilon({\bf k}) = \frac{\hbar^{2} {\bf k}^{2}}{2 M^{*}} - e \phi({\bf x})
,
\end{equation}
so that the local distribution function reads
\begin{equation}
n(\bf x) = \int \frac{d {\bf k}}{(2 \pi)^{2}}
\frac{1}{1 + \exp{\{\beta(\hbar^{2} {\bf k}^{2}/2 M^{*}
- e \phi({\bf x}) - \mu)\}}}
,
\end{equation}
and the density of the background positive charge is
\begin{equation}
n^{0}(\mu) = \int \frac{d {\bf k}}{(2 \pi)^{2}}
\frac{1}{1 + \exp{\{\beta(\hbar^{2} {\bf k}^{2}/2 M^{*}-\mu)\}}}
.
\end{equation}
Thus, we can write the induced density of electrons as
$\rho^{ind}({\bf x}) = -e\{ n^{0}(\mu + e \phi({\bf x}))
- n^{0}(\mu)\}$.
If $\phi$ is small,
$\rho^{ind}({\bf x}) = - e^{2} \partial n^{0}/\partial \mu
\phi({\bf x})$,
so that
\begin{equation}
\chi({\bf q}) = - e^{2} \frac{\partial n^{0}}{\partial \mu}
= - \frac{M^{*} e^{2}}{2 \pi \hbar^{2}}
.
\end{equation}
The dielectric constant reads
$\epsilon({\bf q}) = 1 + q_{T F}/|{\bf q}|$,
where $q_{T F}$ is the Thomas-Fermi wave vector
\begin{equation}
q_{T F} = \frac{M^{*}}{\hbar^{2}}
\frac{e^{2}}{4 \pi \epsilon_0 \epsilon_r}
.
\label{eq15}
\end{equation}
If one adds an extra charge $Q$ at the origin,
\begin{equation}
\phi({\bf q}) = \frac{1}{\epsilon({\bf q})} \phi^{ext}({\bf q})
= \frac{Q}{2 \epsilon_0} \frac{1}{|{\bf q}| + q_{T F}}.
\end{equation}
A Fourier transform yields
\begin{equation}
\phi({\bf x}) = \frac{Q}{4 \pi \epsilon_0 |{\bf x}|} F(q_{T F} |{\bf x}|)
,
\end{equation}
with
\begin{equation}
F(z) = \int_{0}^{+ \infty} J_0(u) \frac{u}{u+z} du
.
\end{equation}
The interaction is screened if $x \gg 2 \pi/q_{T F}$.
The numerical value of the Thomas-Fermi length $\lambda_{T F}=2 \pi/q_{T F}$
is $\lambda_{T F}=15.6$nm.

\subsection{Linear response approach}
The aim of the section is to present the real space linear response
formalism. From a numerical point of view, this means that we must
discretize the radial coordinate  at a scale smaller than the Thomas-Fermi
length $\lambda_{T F}$.  Within the linear response, the density variation
is linearly related to the local potential, but in a non local way:
\begin{equation}
\label{eq24}
\delta \rho({\bf x}) = \int_{0}^{L} \chi^{(0)} (x,x') V_{loc}(x') dx'
,
\end{equation}
where $x \in [0,L]$ is the radial coordinate , with the origin taken
at the interior edge of the ring. Notice that the conservation of charge
carriers implies that
\begin{equation}
\int_{0}^{L} \chi^{(0)}(x,x') dx = 0
.
\end{equation}
The local potential is the sum of the potential created by the
electrostatic field of the central charge $V_{ext}(x)$,
plus the potential $V_{ind}(x)$ induced by the electrons on the ring.
Since the distance $L$ between the two edges is small ($L = 0.1 \mu m$),
we linearize the Coulomb potential created by the central charge
in the vicinity of the ring, which leads to the following expression
of $V_{ext}(x)$:
\begin{equation}
V_{ext}(x) = \frac{Q e^{2}}{4 \pi \epsilon_0 \epsilon_r (r_0 + L/2)^{2}}
(x-\frac{L}{2})
\end{equation}
On the other hand, the induced potential has the expression
\begin{eqnarray}
V_{ind}(x) &=& \frac{e^{2}}{4 \pi \epsilon_0 \epsilon_r} 
\int_{0}^{L}
(1+\frac{x'}{r_0}) \delta \rho(x') dx'\\
&& \int_{-\pi}^{\pi} \frac{d \theta}{ \sqrt{(1 + \frac{x}{r_0})^{2}
+ (1+\frac{x'}{r_0})^{2} - 2 (1 + \frac{x}{r_0}) (1 + \frac{x'}{r_0})
\cos{\theta}} }
.
\end{eqnarray}
Using the notations $u=1+x/r_0$ and
$v=1+x'/r_0$, the angular integral takes
the form of an elliptic integral
\begin{equation}
\int_{-\pi}^{\pi} \frac{d \theta}
{\sqrt{u^{2}+v^{2}-2 u v \cos{\theta}}}
= 4 \int_{0}^{\pi/2} \frac{d \theta}
{\sqrt{((u+v)^{2}- 4 u v \cos^{2}{\theta}}}\\
=\frac{4}{u+v} K\left(\frac{2 \sqrt{u v}}{u+v}\right)
.
\end{equation}
Since $u$ and $v$ are close to unity, the elliptic integral can be approximated
as
\begin{equation}
K\left(\frac{2 \sqrt{u v}}{u+v} \right) = \ln{\left(
\frac{4 |u+v|}{|u-v|}\right)} + O\left((u-v)^{2}\ln{|u-v|}\right)
,
\end{equation}
which leads to
\begin{equation}
V_{ind}(x)  \simeq \frac{e^{2}}{2 \pi \epsilon_0 \epsilon_r}
\int_0^{L} \ln{\left(\frac{8 r_0}{|x-x'|}\right)} \delta \rho(x') dx'
\label{eq25}
\end{equation}
Notice that the factor $8 r_0$ does not come into account since
\begin{equation}
\int_0^{L} \delta \rho(x) dx = 0.
\end{equation}
The equation (\ref{eq24}) reads
\begin{equation}
\label{eq26}
\delta \rho = \chi^{(0)}(V_{ext}+ V_{ind} \delta \rho)
,
\end{equation}
where the quantities are understood as matrices for $\chi^{0}$ and
$V_{ind}$ and vectors for $\delta \rho$ and $V_{loc}$. A discretization
of the radial coordinate has been assumed, and
we used a parabolic approximation for the integrals.
A special attention has to be paid to the logarithmic divergence of
(\ref{eq25}) which has to be integrated explicitly using a parabolic
approximation. Equation (\ref{eq26}) can be inverted into the form
\begin{equation}
\delta \rho = \frac{1}{1 - \chi^{(0)} V_{ind}} \chi^{(0)} V_{ext}
\label{eq30}
,
\end{equation} 
which has the well known form of a R.P.A. resummation.
Numerically, we use a Gauss-Jordan method for inverting the linear
system (\ref{eq25}). The size of the matrix to be inverted is
about 1000, which corresponds to 1000 points for the discretization
of the interval $[0,L]$. Since $L \sim 0.1 \mu m$, the condition
$L/1000 \ll \lambda_{T F}$ is well respected.
We now examine successively two levels of approximation for the response function
$\chi^{(0)}(x,x')$.

\subsection{Thomas-Fermi approach to the response function}
In the Thomas-Fermi approach, the response function is purely local:
\begin{equation}
\chi^{(0)}_{T F}(x,x') = - \frac{m}{2 \pi \hbar^{2}}(\delta(x-x')-\frac{1}{a})
\end{equation}
Using this form of the response function, we inverted the system (\ref{eq26}).
The resulting density profile is given on figure \ref{Fig4}.
In terms of the effective flux through the sample (flux of the magnetic
field plus flux of the statistical gauge field), one is interested in
\begin{equation}
\phi(x) = - 4 \pi \phi_0 (r_0+L/2) \int_{0}^{x} \delta \rho(x) dx
.
\end{equation}
The function $\phi(x)$ is plotted on figure \ref{Fig5}.
As we shall see in section 5.5, the total
current is given in terms of the mean flux through the ring
\begin{equation}
\overline{\phi} = \frac{1}{L} \int_0^{L}\phi(x) dx
\label{eq31}
\end{equation}
We find that $\overline{\phi}$ is such that a charge $Q=14 e^{-}$ is required
to induce one flux quantum on average through the ring, which means that
the periodicity of the currents as a function of the central charge
is 14 electrons within the infinite square well potential and the
Thomas Fermi approximation.

\subsection{Screening of the central charge with the full response function}
\label{fullresponse}
We calculate the full response function using the first order perturbation
theory in the potential induced by the central charge.We also use the
simplified geometry. The matrix elements of the local potential
$V_{loc}$ on the basis of function (\ref{eq27}) are
\begin{equation}
V_{n,n'}^{loc} = \frac{2}{L} \int_0^{L}
\sin{\left(\frac{\pi}{L}n x \right)}
V_{loc}(x)
\sin{\left( \frac{\pi}{L}n'x\right)} dx
,
\end{equation}
so that
\begin{equation}
\hat{V}_{loc} = \sum_m \sum_{n,n'} V_{n,n'}^{loc}
\psi_{m,n}^{+} \psi_{m,n'}
.
\end{equation}
The first order corrections to the state $|\psi_0\rangle$
of equation (\ref{eq23}) are
\begin{equation}
|\psi\rangle = |\psi_0\rangle +
\sum_m \sum_{n,n'} \frac{V_{n,n'}^{loc}}{E(m,n')-E(m,n)}
\psi_{m,n}^{+} \psi_{m,n'} |\psi_0 \rangle
.
\end{equation}
The second-quantized form of the electron density operator is
\begin{equation}
\hat{\rho}(x,y) = \frac{2}{R L}
\sum_{m,m'} \sum_{n,n'}
e^{- i(m'-m) \frac{2 \pi}{R}x}
\sin{\left( \frac{\pi}{L} n y \right)}
\sin{\left( \frac{\pi}{L} n' y \right)}
\psi_{m,n}^{+} \psi_{m',n'} |\psi_0 \rangle
.
\end{equation}
One can readily calculate the average of $\hat{\rho}(x,y)$ in the presence
of the central charge, which leads to
\begin{eqnarray}
\nonumber
\langle \hat{\rho}(x,y) \rangle_{Q} -
\langle \hat{\rho}(x,y) \rangle_{Q=0} &=&
\frac{4}{R L} \sum_m \sum_{n,n'}
\frac{V_{n,n'}}{E(m,n')-E(m,n)}
n^{0}(m,n')\\
&&
(1-n^{0}(m,n))
\sin{\left(\frac{\pi}{L}n y\right)}
\sin{\left(\frac{\pi}{L}n' y\right)}
,
\end{eqnarray}
where $n^{0}(n,m)$ is unity if $(m,n)$ belongs to the Fermi sea
and zero otherwise. The summation over $m$ can be
evaluated by noticing that the Fermi sea is made up of a collection
of nested one dimensional Fermi seas (one for each channel),
and we obtain
\begin{eqnarray}
\nonumber
\delta \langle \hat{\rho}(x,y)\rangle = \frac{8 M^{*}}{\pi^{2}
\hbar^{2}}
&\{& \sum_{n=1}^{N} \sum_{n'=n+1}^{N}
\frac{\sqrt{\nu^{2}-n^{2}}-\sqrt{\nu^{2}-n'^{2}}}{n^{2}-n'^{2}}
\sin{\left(\frac{\pi}{L} n y\right)}
\sin{\left(\frac{\pi}{L} n' y\right)}
V_{n,n'}^{loc}\\
\label{eq28}
&&
+ \sum_{n=1}^{N} \sum_{n'=N+1}^{+ \infty}
\frac{\sqrt{\nu^{2}-n^{2}}}{n^{2}-n'^{2}}
\sin{\left(\frac{\pi}{L} n y\right)}
\sin{\left(\frac{\pi}{L} n' y\right)}
V_{n,n'}^{loc}
\}
.
\end{eqnarray}
In this expression, $\nu$ is related to the Fermi energy by
\begin{equation}
E_F = \frac{\hbar^{2}}{2 M^{*}} \frac{\pi^{2}}{L^{2}} \nu^{2}
\end{equation}
and $N$ is the integer such as $\nu \in [N,N+1[$. We deduce from
(\ref{eq28}) the response function $\chi^{(0)}(y,y')$:
\begin{equation}
\nonumber
\chi^{(0)} = \frac{16 M^{*}}{\pi^{2}\hbar^{2}L} \left\{
 \sum_{n=1}^{N} \sum_{n'=n+1}^{N}
\frac{\sqrt{\nu^{2}-n^{2}}-\sqrt{\nu^{2}-n'^{2}}}{n^{2}-n'^{2}}
\Lambda_{n,n'}^{(ISW)}(y.y')
+
\sum_{n=1}^{N} \sum_{n'=N+1}^{+ \infty}
\frac{\sqrt{\nu^{2}-n^{2}}}{n^{2}-n'^{2}}
\Lambda_{n,n'}^{(ISW)}(y,y')\right\}
\label{eq29}
,
\end{equation}
with
\begin{equation}
\Lambda_{n,n'}^{(ISW)}(y,y') = \sin{\left(\frac{\pi}{L} n y \right)}
\sin{\left(\frac{\pi}{L}n' y \right)}
\sin{\left(\frac{\pi}{L} n y' \right)}
\sin{\left(\frac{\pi}{L} n' y' \right)}
.
\end{equation}
The symbol IFW stands for "infinite square well".
With the form (\ref{eq29}) for $\chi^{(0)}$,
it is straighforward to compute the
density of electrons, using the matrix relation (\ref{eq30}).
The profile of the electron density variations induced by the central
charge is plotted on figure \ref{Fig6}. We notice the existence of 
oscillations at the Thomas-Fermi wave vector, which were absent in the
calculation using the Thomas-Fermi approach to the response function.
However, these oscillations do not
affect too much the averaged quantities under
interest. The total flux through the sample $\phi(x)$ is plotted on figure
\ref{Fig7}, and has the same shape as the flux computed in the
Thomas-Fermi approximation. The mean flux through the ring is such as
18.2 $e^{-}$  are required to produce one flux quantum through
the ring within the infinite square well approximation and within
a calculation including the full response function.

\subsection{Electrons in a parabolic potential well}
In experiments, the edges of the ring may not be so well--defined,
and the electronic density will exhibit a gradual fall off
to zero. This section is devoted to model the effects
of ill--defined edges.
To do so, we use the same techniques
as before, and use a parabolic confining well instead of
an infinitely deep well. Within the mean field treatment,
the Hamiltonian of the quasiparticles reads
\begin{equation}
H = \frac{{\hat p}^{2}}{2 M^{*}} + \frac{1}{2}
M^{*} \omega^{2} y^{2}
.
\end{equation}
As before, we start from a Slater determinant (\ref{eq23}),
with one electron wave functions
\begin{equation}
\varphi_{m,n}(x,y) = \frac{1}{\sqrt{R}}
\left( \frac{\beta^{2}}{\pi} \right)^{1/4}
\frac{1}{\sqrt{2^{n} n!}} \exp{\left( - \frac{\beta^{2}
y^{2}}{2} \right)} H_n(\beta y) e^{-i m \frac{2 \pi}{R} x}
\end{equation}
where the inverse length scale $\beta$ is
\begin{equation}
\beta = \sqrt{\frac{M^{*} \omega}{\hbar}}
\end{equation}
and $H_n$ are the Hermite polynomials.
The radial density of the Slater determinant is plotted
on figure \ref{Fig8}, where the strength of the harmonic
potential has been adjusted such as $70 \%$ of the
electronic density lies within the one micrometer wide ring.
Friedel oscillations are visible. From a computational
point of view, the Hermite polynomials may reach
huge values. It is thus more practical to calculate
recursively
$\tilde{H}_n= H_n / \sqrt{2^{n} n!}$, with the
following recursion relations
\begin{equation}
\tilde{H}_{n+1}(u) = \sqrt{\frac{2}{n+1}} u
\tilde{H}_n(u)
- \sqrt{\frac{n}{n+1}} \tilde{H}_{n-1}(u)
,
\end{equation}
and $\tilde{H}_0(u) = 1$, $\tilde{H}_1(u) = \sqrt{2} u$.
The same approach as in section \ref{fullresponse} can
be carried out and the response function is
\begin{equation}
\chi^{(0)} = \frac{2 \sqrt{2}}{\pi^{2}}
\frac{M^{*} \beta}{\hbar^{2}}
e^{- \beta(y^{2}+y'^{2})}
\left\{ \sum_{n=0}^{N} \sum_{n'=n+1}^{N}
\frac{\sqrt{\nu - n} - \sqrt{\nu - n'}}{n-n'}
\Lambda_{n,n'}^{(HW)}(y,y')
+ \sum_{n=0}^{N} \sum_{n'=N+1}^{+ \infty}
\frac{\sqrt{\nu-n}}{n-n'} \Lambda_{n,n'}^{(HW)}(y,y')
\right\}
,
\end{equation}
with
\begin{equation}
\Lambda_{n,n'}(y,y') = \tilde{H}_n(\beta y) \tilde{H}_{n'}(\beta y)
\tilde{H}_n(\beta y') \tilde{H}_{n'}(\beta y')
,
\end{equation}
and $\nu$ is defined by $E_F = (\nu+1/2)\hbar \omega$.
Using similar techniques as in section \ref{fullresponse},
we get the radial density variations (figure \ref{Fig9})
as well as the induced flux (figure \ref{Fig10}).
We come to the conclusion that a central charge
approximately equal
to 9 electrons is required to generate one flux
quantum through the ring, which suggests that the
Fermi sea in a parabolic well is more sensitive to the
central charge as in the infinite square well. This is due
to the fact that the fraction of the electronic density
outside the $0.1 \mu m$ ring experience is closer
to the central charge (as far as the inner edge is
concerned) and thus experiences a stronger potential.
We conclude to the importance of whether there
is a sharp edge or a graduate fall--off to zero of the
electronic density. In the former case, one flux quantum
could be generated through the ring with a smaller charge
than in the latter case.

\subsection{Currents induced by the charge}
The magnetic flux through the ring generates permanents currents, in the same
way as a flux through the hole of the ring generates currents. Let us call
${\bf \delta A}$ the variation in the vector potential due to the presence of 
the charge. The variation in the Hamiltonian due to the shift in the vector
potential is
\begin{equation}
\delta H = \frac{e^{2}}{2 M^{*}} (2 {\bf A}.{\bf \delta A} +
{\bf \delta A}^{2})
+ \frac{i e \hbar}{M} {\bf \delta A}.{\bf \nabla}
.
\label{eq21}
\end{equation}
Applying first order perturbation theory, we obtain the variation of a given
energy level
\begin{equation}
\langle \delta H \rangle = \frac{e^{2}}{M^{*}}
 \langle {\bf A}.{\bf \delta A} \rangle
-\frac{e \hbar}{M^{*}} m \langle \frac{\delta A(r)}{r} \rangle
.
\end{equation}
We now make the approximation that the spatial dependence of ${\bf \delta A}$
and $\delta A(r) / r$ are averaged on the ring, that is to say
\begin{equation}
\frac{\delta A(r)}{r} \simeq \frac{\overline{\phi}}{2 \pi r_0^{2}}
,
\end{equation}
where $\overline{\phi}$ is the average magnetic flux induced by the central electric
charge.
The first term in $\delta H$ is simply an energy variation independent on $n$
and $m$. This term is then dropped. Thus we get
\begin{equation}
\langle \delta H \rangle = - \frac{e \hbar}{M^{*}} m \frac{\overline{\phi}}
{2 \pi r_0^{2}},
\end{equation}
so that the total energy reads
\begin{equation}
E(m,n) = \frac{\hbar^{2}}{2 M^{*} r_0^{2}}
(m - \frac{\overline{\phi}}{\phi_0})^{2} -\frac{\hbar^{2}}{2 M r_0^{2}}
(\frac{\overline{\phi}}{\phi_0})^{2}
.
\end{equation}
If we apply the result of section 4 about Aharonov Bohm
currents, we find that the currents
are periodic in $\overline{\phi}$. The maximum value of the current is given by
(\ref{eq16}) and the period is such as $\overline{\phi}=\phi_0$.

\section{Numerical approach in the spherical geometry}
We now make use of the spherical geometry in order to perform numerical
computations with a small number of electrons on a sphere of radius $R$.
A magnetic monopole is put at the center of the sphere, creating a total
magnetic flux through the sphere equal to $2 S \phi_0$. The Dirac's
monopole quantization requires $2 S$ to be an integer. An electric
charge $Q$ is put at the north pole of the sphere. The extra charge is
treated as a classical, point charge. We first generalize the mean field
theory argument to the spherical geometry case. In a second step, we present
numerical computations for a small number of electrons on the sphere for
various filling fractions.

\subsection{Mean field theory at $\nu=1/2$}
We adapt the argument of the mean field theory, previously established
in the geometry of the ring, to the spherical geometry.
The filling fraction
\begin{equation}
\nu = \frac{N-1}{2 S}
\end{equation}
is chosen to be $1/2$ in this section. The non local gauge transformation
leads to a statistical field which generates a flux equal to
$- 2 \phi_0(N-1)$ because there is no statistical flux coming from
a particle onto itself.
The mean field Hamiltonian simply corresponds to fermions on the sphere
in the absence of a magnetic field. If one neglects the Coulomb
interactions, the Hamiltonian is diagonal on the basis of the spherical
harmonics $Y_{l,m}(\theta,\varphi)$ normalized such as
\begin{equation}
\int Y_{l,m}(\theta,\varphi) d \Omega= 1
\end{equation}
The wave function is simply
$\psi_{l,m}(\theta,\varphi) = Y_{l,m}(\theta,\varphi) / R$. The single
particle states are labelled by the positive integer $l$ and the integer $m$
such as $-\l \le m \le l$. The energy of a state labelled by $(l,m)$ reads
\begin{equation}
E(l,m) = \frac{\hbar^{2}}{2 M^{*} R^{2}} l(l+1)
. 
\end{equation}
In the presence of a negative (positive) electric charge at the north pole,
a depletion (accumulation) of electrons arises at the north pole,
and an accumulation (depletion) arises at the south pole.
The mean field model in the presence of the charge at the north pole
consists of electrons in a zero magnetic field plus a flux tube
$\overline{\phi}$ penetrating through the south pole and emerging
at the north pole. The problem of quantum motion around a flux tube
$\overline{\phi}=\alpha \phi_0$ in the spherical geometry was considered
in \cite{Ref15}. We now rederive the solution.
This problem is non perturbative in $\overline{\phi}$, since infinite
quantities
appear in the first order perturbation theory in $\overline{\phi}$.
This is why we adopt an algebraic approach.
The flux tube $\overline{\phi}$
is absorbed in a gauge transformation, leading to multivalued wave functions
\begin{equation}
\psi(r,\theta+2 \pi)= e^{2 i \pi \gamma} \psi(r,\theta)
.
\end{equation}
The eigenvalues of $l_z$ are thus quantized by $l_z = m + 2 \pi \gamma$,
$m$ being an integer.
The eigenstates of the Hamiltonian are also eigenstates of ${\bf l}^{2}$,
since
\begin{equation}
H = - \frac{\hbar^{2}}{2 m} {\bf \Delta}
= - \frac{\hbar^{2}}{2 M^{*} R^{2}}{\bf l}^{2}
.
\end{equation}
A priori, it is not obvious to produce a basis of commun eigenvectors
to ${\bf l}^{2}$ and $l_z$., since the rotational invariance is broken
by the presence of the flux tube. Nonetheless, as we shall see, it is
possible to diagonalize simultaneously ${\bf l}^{2}$ and $l_z$.
The spherical representation of the kinetic momentum algebra is \cite{Ref16}:
\begin{eqnarray}
l^{+} &=& e^{i \varphi} \left( \frac{\partial}{\partial \theta}
+ i \cot{\theta} \frac{\partial}{\partial \varphi} \right)\\
l^{-} &=& e^{-i \varphi} \left( - \frac{\partial}{\partial \theta}
+ i \cot{\theta} \frac{\partial}{\partial \varphi} \right)\\
l_z &=& \frac{1}{2} [l^{+},l^{-}] = \frac{1}{i} \frac{\partial}
{\partial \varphi}
\end{eqnarray}
The operator algebra gives rise to two ladders of states. The $(-)$ ladder
corresponds to states descending, by the repeated action of $l^{-}$,
from the highest weight state $\psi_0^{(-)}$, such as $l^{+} \psi_0^{(-)}=0$.
The states $\psi_0^{(-)}$ have the form
\begin{equation}
\psi_0^{(-)}(\theta,\varphi) = f^{(-)}(\theta) e^{i(l+\gamma) \varphi}
,
\end{equation}
with $l$ an integer, and is such as
\begin{equation}
\left( \frac{\partial}{\partial \theta} + i \cot{\theta}
\frac{\partial}{\partial \varphi} \right)
e^{i(l+\gamma) \varphi} f^{(-)}(\theta) = 0
,
\end{equation}
so that $f^{(-)}(\theta) \propto \left( \sin{\theta} \right)^{l+\gamma}$,
with $l+\gamma > 0$. We take $\gamma \in ]0,1[$, and $l = 0,1,...$ For
a given $l$, we can produce $l+1$ descending states by the repeated action
of $l^{-}$. For these states, $l_z = \gamma, 1+\gamma,...,l+\gamma$ and
${\bf l}^{2} = (l+\gamma)(l+\gamma +1)$.
The $(+)$ ladder correspond to states ascending, by the repeated
action of $l^{+}$, from the highest weight state $\psi_0^{(+)}$
such as $l^{-} \psi_0^{(+)} = 0$. $\psi_0^{(+)}$ has the form
\begin{equation}
\psi_0^{(+)}(\theta,\varphi) =
f^{(+)}(\theta) e^{i(l+\gamma) \varphi}
,
\end{equation}
with $l$ an integer, and is such as
\begin{equation}
\left( - \frac{\partial}{\partial \theta} + i \cot{\theta} \frac{\partial}
{\partial \varphi} \right) e^{i(l+\gamma) \varphi} f^{(+)}(\theta) = 0
,
\end{equation}
so that $f^{(+)}(\theta) \propto (\sin{\theta})^{-(l+\gamma)}$, with
$l+\gamma<0$, so that $|l| \le -1$.
The repeated action of $l^{+}$ produces $|l|$ states with
$l_z = l + \gamma , l + \gamma + 1,...,-1+ \gamma$ and ${\bf l}^{2}=
(l+\gamma)(l-1+\gamma)$.

We now enumerate the first states. With $l=0$, $E=\gamma(\gamma+1)$ and the
degeneracy $g$ is $1$. With $l=-1$, $E=(1-\gamma)(2-\gamma)$ and $g=1$.
With $l=1$, $E=(1+\gamma)(2+\gamma)$ and $g=2$.
If $l=-2$, $E=(2-\gamma)(3-\gamma)$ and $g=2$.
If $l=2$, $E=(2+\gamma)(3+\gamma)$ and $g=3$.
If $l=-p$, $E=(p-\gamma)(p+1-\gamma)$ and $g=p$ and if
$l=p$, $E=(p+\gamma)(p+1+\gamma)$ and $g=p+1$.

We now turn to the calculation of permanent currents, since this is
the quantity we shall compute using numerical diagonalizations for a
small number of electrons.
We distinguish between two cases: $p(p-1) \le N \le p^{2}$
and $p^{2} \le N \le p(p+1)$. First, if $p(p-1) \le N \le p^{2}$,
we note $\lambda = N-p(p-1)$. The total energy is given by
\begin{eqnarray}
E_{tot} &=& \sum_{m=1}^{p-1} \left( (m-1+|\gamma|)(m+|\gamma|)
+(m-|\gamma|)(m+1-|\gamma|) \right) m + (p-1+|\gamma|)(p+|\gamma|) \lambda\\
&=& \frac{1}{2} p^{2}(p-1)^{2} + p(p-1) \lambda
+ \left( (2p-1) \lambda - p(p-1) \right) |\gamma| + \left( \lambda + p(p-1)
\right) \gamma^{2}
\end{eqnarray}
for $\gamma \in [-1/2,1/2]$.
If $p^{2} \le N \le p(p+1)$, we note $\lambda'=N-p^{2}$ and, for
$\gamma \in [-1/2,1/2]$, we obtain
\begin{equation}
E_{tot} = \frac{1}{2} p^{2}(p-1)^{2} + p^{2}(p-1) + p(p+1) \lambda'
+ \left( p^{2} -(2 p +1) \lambda' \right) |\gamma|
+ \left( p^{2} + \lambda' \right) \gamma^{2}
.
\end{equation}
The current $d E_{tot}/d \gamma$ is discontinuous for half integer
values of $\gamma$, as for the ring, but further discontinuities
appear for integer values of $\gamma$.
The jump in the current at $\gamma=0$ is
\begin{equation}
\frac{d E_{tot}}{d \gamma}(0^{+})
- \frac{d E_{tot}}{d \gamma}(0^{-}) = 2 \left( (2 p -1) \lambda -p(p-1)
\right)
,
\end{equation}
or
\begin{equation}
\frac{d E_{tot}}{d \gamma}(0^{+})
- \frac{d E_{tot}}{d \gamma}(0^{-})
= 2 \left( p^{2} - (2p+1) \lambda' \right)
.
\end{equation}
The first term corresponds to the case $p(p-1) \le N \le p^{2}$,
and the second case to $p^{2} \le N \le p(p+1)$. The jump of the current at
$\gamma=1/2$ is
\begin{equation}
\frac{d E_{tot}}{d \gamma}(\frac{1}{2}^{+})
- \frac{d E_{tot}}{d \gamma}(\frac{1}{2}^{-}) =
- 4 p \lambda
,
\end{equation}
or
\begin{equation}
\frac{d E_{tot}}{d \gamma}(\frac{1}{2}^{+})
- \frac{d E_{tot}}{d \gamma}(\frac{1}{2}^{-}) = - 4 p(p-\lambda')
.
\end{equation}

\subsection{Numerical procedure}
We assume that the magnetic field is large enough to neglect the excitations
from the first Landau level to higher Landau levels.
We propose to
diagonalize numerically the Coulomb interaction inside the Hilbert space
generated by the set of the occupations of the lowest Landau level.
The quantum Hall effect on the sphere has been studied by Haldane
\cite{Ref14}. He proposes a set of coherent states which span the
lowest Landau level Hilbert space. From this set of coherent states, we
can extract the following basis of the lowest Landau level Hilbert space
\begin{equation}
\phi_{\alpha}({\bf x}) = \langle {\bf x}|\alpha \rangle =
N_{\alpha} (\sin{\frac{\theta}{2}})^{\alpha}
(\cos{\frac{\theta}{2}})^{2 S - \alpha}
e^{i \varphi(S-\alpha)}
\label{eq5}
,
\end{equation}
where $N_{\alpha}$ is chosen in such a way that
\begin{equation}
\int|\phi_{\alpha}({\bf x})|^{2} d {\bf x} =1
.
\end{equation}
The label $\alpha$ indexes the orbitals of the lowest Landau level and
runs from $0$ to $2 S$.
One needs to compute the matrix elements of the Coulomb interaction.
The interaction $V_1$ between the electrons on the sphere and the classical
charge located at the north pole is a one-body operator and the interactions
$V_2$ between the electrons on the sphere is represented by a two-body operator.
Using the second quantization, it is straightforward to derive expressions
for the matrix elements of $\hat{V}_1$ and $\hat{V}_2$ on the basis of the
states with all the possible occupation numbers of the lowest Landau level.
We do not give here the details of the calculations. We used the
the results of \cite{Fano} relating the matrix elements
of the Coulomb potential to Clebsh--Gordan coefficients.
One is left with a symmetric
matrix to be diagonalized using a Lanczos method in each sector
of the $z$ component of the angular momentum (since the Hamiltonian
is invariant under any rotation around the $z$ axis).
The ground state is used to
calculate the expectation value of the one-body current operator.
The total intensity is obtained after an integration of the current
$j(\theta)$ over the angular coordinate $\theta$.

\subsection{Results}
We diagonalized the Coulomb interaction at half filling
with $N=8$ electrons
(figure \ref{Fig12}) and $N=9$ electrons
(figure \ref{Fig13}). Obviously, the intensity is not
periodic as the central charge varies. This is due to
the fact that only a small number of electrons is present
in our numerical diagonalizations. As the central charge
is tuned, level crossing occurs, namely the $z$ component
of the angular momentum of the ground state is varied
as the central charge is varied. Notice that these level
crossings are allowed because they correspond to crossing
of levels in different sectors of the Hilbert space.
Due to the small number of electrons, as the central classical
charge is increased, one reaches a point where no more crossing
occurs since the Hilbert space, restricted
to the lowest Landau level is finite. Above this
point, the intensity saturates. If the central charge
is large and positive, the electrons are attracted at the north
pole and the current is thus positive. By contrast, if
the classical charge is large and negative, the electrons
are repelled from the north pole, and mainly located on
the south hemisphere, and the current is negative
(see figure \ref{Fig14}). However, if the classical
charge is not too important, we observe discontinuous
variations of the intensity as a function of the
central charge, which is very much reminiscent of the
previously calculated mean field behavior
even though we were not able to related the position
of the discontinuities to the ones of the
mean field model.
We checked that the discontinuities in the current
correspond to level crossings at the bottom of the
spectrum in the sense that each jump in the current
is associated with the fact that the $z$ component
of the angular momentum of the ground state changes.
We conclude that
the mean field scnario for the intensity as a function
of the central charge is qualitatively supported by our
numerical calculations. At the quantitative
level, it seems rather difficult to conclude since
we can only carry out numerical diagonalizations for
a small number of electrons. In particular, we
cannot conclude on the existence of periodic variations
of the intensity as a function of the classical charge
since the available sizes are too small.

\section{Conclusion}
We have shown that a charge in the middle of ring of $\nu=1/2$ electrons
induces currents which vary periodically as a function of the central
charge. The mechanism for the generation of these periodic currents
involves a polarization of the electron liquid on the ring.
If the extra charge is positive,
a positive density variation appears
on the interior edge of the ring, and a depletion of negative charges
appears on the external edge of the sample. These charges screen the
field of the central charge.
The appearance of currents on the ring is mediated by the variation of the
Chern-Simons gauge field at mean field level. The presence of polarization
charges on the edges on the sample induces a non-zero average flux
through the ring. By increasing the charge, one should be able to produce
periodic currents in the ring. The order of magnitude of these Aharonov-Bohm
currents is such that they can be measured \cite{Ref18}.
The amplitude of the current fluctuations is typically 2.0 nA for a ring
of radius 1 $\mu m$ and transverse dimension 0.1 $\mu m$ and a magnetic
field of 20 T.
If one can produce a continuous charge
in the middle of the ring, one should be able to find experimentally the
period of the phenomenon. Our mean field calculations show
that the period is of the order of 14 electrons (calculated with
the full response function) with an infinite square potential.
However, the periodicity of the phenomenon is sensitive to
the existence of a graduate density fall--off at the edges, as
is expected to occur in experiments. In the case of a parabolic
well with $30\%$ of the electronic density outside the $0.1$
micrometer ring, we found a $9$ electrons periodicity.
Residual currents should also appear in the absence of the charge, because
of the presence of a magnetic flux through the ring.
The experimental observation of such currents should be a direct test of
the existence of a statistical gauge field in the $\nu=1/2$ quantum Hall
effect.
We also carried out numerical computations in the spherical geometry.
The mean field picture at $\nu=1/2$ predicts discontinuities in the intensity
plotted as a function of the extra charge and also
the existence of periodic variations.
In our simulations, the current is not periodic as a function of the
charge at the north pole, because of a too small number of electrons.
However, we find the existence of discontinuities, reminiscent
of the predictions of the mean field treatment. We conclude to the
qualitative
consistency between our numerical diagonalizations and
the mean field theory.

Our analysis shows that the type of experiment we consider would be
a good test for the mean field theory of the $\nu=1/2$ quantum
Hall effect since the presence of an extra charge allows to explore
excited states properties, and this in a way which goes beyond linear
response theory, since it involves level crossings and non-trivial
reshuffling of the many body ground state as the external charge
is varied.

Finally, the case $\nu=1/3$ would also be
an interesting case to examine and the physics might
be quite different from the $\nu=1/2$ case investigated
here. It would be especially
interesting to understand the interplay between uncompressible
bulk excitations and compressible edge states \cite{Wen} as the
central charge is varied. This question will be addressed in
a future work.

The authors acknowledge L. L\'evy for stimulating discussions which
encouraged us to carry on this investigation.
R.M. acknowledges J.C. Angl\`es d'Auriac and F.V. de Abreu
for assistance with the numerical work.

\newpage
\renewcommand\textfraction{0}
\renewcommand\floatpagefraction{0}
\noindent {\bf Figure captions}

\begin{figure}[h]
\caption{}
\label{Fig1}
(a): annulus geometry. (b): simplified geometry.
\end{figure}

\begin{figure}[h]
\caption{}
\label{Fig2}
Fermi sea of quasiparticles at the mean field level
in the simplified geometry. $k_x$ is a multiple of $2 \pi/R$
and $k_y$ is a multiple of $\pi / L$.
\end{figure}

\begin{figure}[h]
\caption{}
\label{Fig3}
Electronic density of the Slater determinant
(\ref{eq23}) $\rho(x) = | \langle x | \psi_0
\rangle | ^{2}$. The electronic density vanishes at the edges and exhibits
Friedel oscillations. The fermi sea contains 6 channels and 1494 electrons.
The density is plotted in
$\mu m^{-2}$ and the radial coordinate in $\mu m$.
The electrons are confined in an infinite square well.
\end{figure}

\begin{figure}[h]
\caption{}
\label{Fig4}
Profile of the electronic density variations induced by the central
charge in the Thomas-Fermi approach to the response function. The computation
has been made in the simplified geometry.
The central charge is $Q=-1 e^{-}$.
The density is plotted in
$\mu m^{-2}$ and the radial coordinate in $\mu m$.
The electrons are confined in an infinite square well.
\end{figure}

\begin{figure}[h]
\caption{}
\label{Fig5}
Flux through the ring in the Thomas-Fermi approach to the response function.
The central charge is $Q=-1 e^{-}$.
The flux is plotted in units of the flux quantum $\phi_0$ and the radial
coordinate in $\mu m$.
The electrons are confined in an infinite square well.
\end{figure}

\begin{figure}[h]
\caption{}
\label{Fig6}
Profile of the electronic density variations induced by the central
charge with the full response function.
The computation
has been made in the simplified geometry.
The central charge is $Q=-1 e^{-}$.
The density is plotted in
$\mu m^{-2}$ and the radial coordinate in $\mu m$.
The electrons are confined in an infinite square well.
\end{figure}

\begin{figure}[h]
\caption{}
\label{Fig7}
Flux through the ring with the full response function.
The central charge is $Q=-1 e^{-}$.
The flux is plotted in units of the flux quantum $\phi_0$ and the radial
coordinate in $\mu m$. The origin of the radial coordinate
is chosen at the inner edge. The unit length is the micrometer
and the density is measured in micrometers$^{-2}$.
The electrons are confined in an infinite square well.
\end{figure}

\begin{figure}[h]
\caption{}
\label{Fig8}
Radial density of the Slater determinant with a parabolic
potential well. The Fermi sea contains 8 channels and the strength
of the parabolic potential has been adjusted such as $70 \%$
of the density is located on the 1 micrometer wide ring.
Friedel oscillations are visible. The radial coordinate
is counted in micrometers from the inner edge. The
density is measured in micrometers$^{-2}$.
\end{figure}

\begin{figure}[h]
\caption{}
\label{Fig9}
Variations of the electronic density induced by
a central charge equal to minus one electron, with a parabolic
confining potential. The origin of distances is
the center of the ring (minimum of the potential).
The distances are measured in micrometers and the
density in micrometers$^{-2}$.
\end{figure}

\begin{figure}[h]
\caption{}
\label{Fig10}
Flux through the ring induced by a central charge equals
to minus one electron with a parabolic confining well.
The distances are measured in micrometers and the densities
are in micrometers$^{-2}$. The origin of the radial coordinate
is the center of the well.
\end{figure}

\begin{figure}[h]
\caption{}
\label{Fig11}
Variations of the electronic current as a function of the central charge
in the mean field approach. The period of the variations depends of
the type of approximation (Thomas Fermi or full response function)
and on the nature of the confining well (see the text).
The residual current at $Q=0$ originates from the
Aharanov-Bohm currents induced by the presence of the flux
in the hole of the ring.
\end{figure}

\begin{figure}[h]
\caption{}
\label{Fig12}
Current as a function of the classical charge
located at the north pole.
The intensity is $I=\int j(\theta) d \theta$. The radius $R$ of the sphere
is taken to be $1$. The charge $-Q$ is plotted in units of the electronic
charge $e$. The current is plotted in units of $2 e \hbar/M^{*}$.
The system contains 8 electrons.
\end{figure}

\begin{figure}[h]
\caption{}
\label{Fig13}
Current as a function of the classical charge
located at the north pole.
The intensity is $I=\int j(\theta) d \theta$. The radius $R$ of the sphere
is taken to be $1$. The charge $-Q$ is plotted in units of the electronic
charge $e$. The current is plotted in units of $2 e \hbar/M^{*}$.
The system contains 9 electrons.
\end{figure}

\begin{figure}[h]
\caption{}
\label{Fig14}
Current circulation on the sphere for a large positive
classical charge (a) and a large negative classical
charge (b). In (a) the electrons gather at the
north pole, due to the attractive Coulomb interaction between
the electron gas and the classical charge. The current in
this case is clockwise. In (b), the electrons
are repelled from the north pole and the circulation
of the current is counterclockwise.
\end{figure}

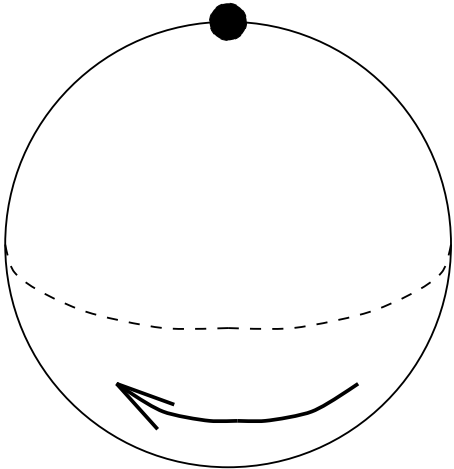
\end{document}